\documentclass[10pt]{article}

\usepackage[left=25mm,right=25mm,top=25mm,bottom=25mm,paper=a4paper]{geometry}
\usepackage{cite} 
\usepackage{hyperref}
\usepackage{amsmath,amssymb}
\usepackage{authblk}
\usepackage{graphicx}
\usepackage[shortlabels]{enumitem}
\usepackage{xcolor}
\usepackage{float} 

\def\be{\begin{equation}} 
\def\ee{\end{equation}}   
\def \eea{\end{eqnarray}}
\def \bea{\begin{eqnarray}}

\newcommand{\puc}{Instituto de F{\'i}sica, Pontificia Cat{\'o}lica Universidad de Chile, Av. Vicu{\~n}a Mackenna 4860, Santiago, Chile}
\newcommand{\vienaTU}{Institut f\"ur Theoretische Physik, TU Wien,
Wiedner Hauptstrasse 8-10, 1040 Vienna, Austria}
\newcommand{\vienaATI}{TU Wien, Atominstitut, Stadionallee 2, 1020 Vienna, Austria}
\newcommand{\cecs}{Centro de Estudios Cient\'{\i}ficos (CECs), Arturo Prat 514, Valdivia, Chile}
\newcommand{\uss}{Universidad San Sebastián, General Lagos 1163, Valdivia, Chile}
\newcommand{\UA}{Departamento de Física Aplicada, Universidad de Alicante, Campus de San Vicente del Raspeig, E-03690 Alicante, Spain.}

\begin{document}
\title{
A Universe from a Lagrangian Fixed Point
}

\author[1,2]{Pedro D. Alvarez}
\affil[1]{\cecs}
\affil[2]{\uss}

\author[3,4,5]{Benjamin Koch 
\thanks{E-mail: \href{mailto:benjamin.koch@tuwien.ac.at}{\nolinkurl{benjamin.koch@tuwien.ac.at}}}}
\affil[3]{\vienaTU}
\affil[4]{\vienaATI}
\affil[5]{\puc}

\author[3,4]{Ali Riahinia
\thanks{E-mail: \href{mailto:ali.riahinia@tuwien.ac.at}{\nolinkurl{ali.riahinia@tuwien.ac.at}}}}

\author[6]{Angel Rincon
\thanks{E-mail: \href{mailto:angel.rincon@ua.es}{\nolinkurl{angel.rincon@ua.es}}}
}
\affil[6]{\UA}

\maketitle

\begin{abstract}
In this paper, we investigate the theoretical possibility that a Lagrangian fixed point, when applied to cosmological models, can drive dynamical evolution towards a bouncing universe. We analyze the physics of a Lagrangian fixed point within the context of a gravitational average effective action featuring scale-dependent couplings. To explore this concept, we develop a toy model set in a four-dimensional, spatially flat spacetime, anchored by a Lagrangian fixed point. Solving the cosmological equations of this model analytically, we identify several non-trivial solution branches. These branches are characterized by a modified scale factor and dynamic gravitational couplings, offering new insights into the behavior of cosmological models under these conditions.
\end{abstract}

\tableofcontents

\section{Introduction and hypothesis}\label{Sec_Intro}



The quest for quantum gravity is vital for our understanding of the fundamental laws that govern the universe and could potentially lead to breakthroughs in areas such as cosmology, particle physics, and our understanding of the nature of reality itself.
One promising approach to quantum gravity is associated with 'Asymptotic Safety,' which posits the existence of an ultraviolet (UV) {{\bf fixed point}}. This theory suggests that gravity becomes well defined at very high energies. This concept could lead to a consistent quantum theory of gravity, help in renormalizing quantum gravity theories, potentially unify fundamental forces, and provide insights into the early universe. However, it remains unclear how these theoretical results might resolve the notorious problems of general relativity, such as the big-bang singularity.

On the other hand, in the context of the study of the early universe, models which are tailored for the description of cosmological evolution such as
bouncing cosmologies are particularly interesting. They propose a scenario in which the traditional Big Bang singularity is avoided. Instead of the universe beginning from an infinitely small and infinitely dense point, bouncing models suggest that the universe underwent a contraction phase prior to its current expansion. This implies that the universe "bounced" back from a highly compressed but finite state, thereby circumventing the physical and mathematical complications associated with a singularity. This approach enriches our understanding of the origins of the universe by sidestepping undefined physics at a singularity. However, in the context of asymptotic safety quantum gravity, the bounce arises naturally from the model as the Null Energy Condition (NEC) is violated in this framework (see Ref.\cite{Ijjas:2016pad,Bhattacharjee:2020eec} and the references therein). Thus, on theoretical grounds certain problems of quantum gravity can be solved by the existence of a {\bf{ fixed point}} in the effective action of quantum gravity, while in practical cosmological models singularities can be avoided with the help of {\bf{ bouncing cosmologies }}.\\ In this paper, we study the existence of a fixed point in the effective average Lagrangian of asymptotic safety quantum gravity where a bounce arises naturally in the model by violating the NEC and its consequences in the cosmology.\\

In this paper and at this stage, our work is still theoretical, focusing on the above research question.
The model does not aspire to provide a realistic detailed model of the evolution of our universe. 
We dedicate the remainder of the Introduction section to clarify our definitions of the concepts of ``effective action'', ``scale invariance'', ``physically meaningful scale-setting'', and ``fixed point of the Lagrangian''. 
In Section
\ref{Sec_Model} we present a toy model that incorporates these concepts. In section \ref{cosmo}, we will focus on the cosmological scenario of this toy model, in particular, in the very early universe, by solving the corresponding scale-dependent Friedman equations with an ELFP.
Our results are discussed and summarized in Section \ref{sec_Discussion}.

\subsection{Effective average action}

The effective average action (EAA hereafter) is a concept that arises in the context of the functional renormalization group approach to quantum field theories, and it has been applied to the study of quantum gravity~(see
\cite{Weinberg:1976xy,Wetterich:1992yh,Morris:1993qb,Bonanno:2000ep,Reuter:2001ag,Litim:2002xm,Reuter:2004nv,Bonanno:2006eu,Niedermaier:2006ns,Percacci:2007sz,Dupuis:2020fhh} and references therein). This tool is particularly valuable for analyzing non-perturbative aspects of quantum field theories. The idea has also motivated some closely related approaches in which classical backgrounds are modified by the inclusion of quantum features, a side effect of considering an effective average action. In the context of black hole physics, the improved black hole formalism \cite{Ishibashi:2021kmf,Ladino:2023zdn,Torres:2017gix,Ladino:2022aja,Rincon:2020iwy}, 
in the context of cosmological solutions, running vacuum models \cite{Sola:2013gha,Shapiro:2003ui,Sola:2005et,Espana-Bonet:2003qjh,Grande:2011xf,Cruz:2023dzn,Panotopoulos:2020kpo}, and in black holes, relativistic stars and cosmology, the scale-dependent formalim \cite{Koch:2016uso,Rincon:2018lyd,Rincon:2018sgd,Rincon:2017goj,Contreras:2019cmf}, among others.

The EAA is a bridge between the classical action of a theory and its quantum effective counterpart~\cite{Wetterich:2001kra}.
%
It contains a scale parameter, often denoted by 
$k$, effective dimensionless couplings $\alpha_i(k)$, and gauge invariant operators ${\mathcal{I}}_i$  with the energy dimension $j(i)+4$. Thus, it can be denoted as infinite sum over these invariants
\be\label{eq_Gammak}
\Gamma_k(\phi)=\int d^4x\;{\mathcal{L}}_k( \phi)
=\int d^4x\sum_i^\infty \alpha_i(k)\cdot {\mathcal{I}}_i(\phi)\cdot k^{j(i)}.
\ee
As the scale $k$ changes, the EAA interpolates between two limits:
\begin{itemize}
    \item At $k \rightarrow  \infty$ The EAA coincides with the bare action of the theory.
    \item At $k \rightarrow  0$ The EAA approaches the full quantum effective action, which incorporates all quantum fluctuations.
\end{itemize}

The idea behind the EAA is to introduce a smooth cut-off scale $k$ that allows for the progressive integration of quantum fluctuations with momenta between zero and $k$. This is opposed to integrating out all quantum fluctuations above $k$, as would typically be the case when deriving the quantum effective action. By doing the integration progressively, one can track how the ``physics'' (e.g., couplings and other parameters) of a quantum field theory changes as more and more quantum fluctuations are included. The mathematical tool for this procedure is the functional renormalization group equation~\cite{Weinberg:1976xy,Wetterich:1992yh,Morris:1993qb,Bonanno:2000ep,Reuter:2001ag,Litim:2002xm,Reuter:2004nv,Bonanno:2006eu,Niedermaier:2006ns,Percacci:2007sz,Dupuis:2020fhh}
\begin{equation}
    \partial_t \Gamma_k = \frac{1}{2} \textbf{STr} \Bigg[ (\Gamma_k^{(2)} + \mathcal{R}_k )^{(-1)} \partial_t \mathcal{R}_k \Bigg].
\end{equation}

When applied to quantum gravity, the effective average action approach shed light into the renormalization group flow of gravitational interactions (see for instance \cite{Buchbinder:1992rb}). This approach has led to insights into the possibility of non-perturbative renormalizability of quantum gravity, often discussed in the context of `asymptotic safety.'' In an asymptotically safe theory of quantum gravity, there exists a UV fixed point where the theory becomes scale-invariant, potentially allowing for a consistent quantum description of gravity without the need for a UV completion or a ``fundamental'' length scale (like the Planck length) where the theory would break down \cite{Benedetti:2009rx,Codello:2008vh,Donkin:2012ud,Lauscher:2001ya,Reuter:1996cp,Niedermaier:2006wt}.

\subsection{Scale dependent gravitational action and equations}

For gravitational theories, the gauge-invariant quantities ${\mathcal{I}}_i$ are typically curvature invariants. 
Retaining only two leading contributions in a low curvature expansion (\ref{eq_Gammak}) reads
\begin{equation}\label{eq_Sk}
 \Gamma_k[g_{\mu \nu}] \equiv 
 \int \mathrm{d}^4x \sqrt{-g}
 {\mathcal{L}}_k[g_{\mu \nu}] + \Gamma_m = \int \mathrm{d}^4 x \sqrt{-g} \Bigg[ \frac{1}{2 \kappa_k} \Bigl(R-2\Lambda_k \Bigl) + \mathcal{L}_m \Bigg],
\end{equation}  
which is know as scale-dependent action in the
Einstein-Hilbert truncation.
Here, $g_{\mu \nu}$ is the metric field, 
$k$ is the renormalization scale, $\mathcal{L}_k$ is the effective Lagrangian density, $g$ is the determinant of the metric field, 
$R$ is the Ricci scalar, $\Lambda_k$ is the cosmological coupling, $\kappa_k \equiv 8\pi G_k$ is Einstein's coupling, and $G_k$ is the Newton's coupling.
Varying (\ref{eq_Sk}) with respect to the metric $g_{\mu \nu}$ yields 
the corresponding
equations of motion are~\cite{Reuter:2004nv,Reuter:2003ca,Koch:2010nn,Domazet:2012tw,Koch:2014joa,Contreras:2016mdt}
\begin{equation}
\label{eq_eomSD1}
G_{\mu \nu} + \Lambda_k g_{\mu \nu} = 8 \pi G_k T_{\mu \nu} -\Delta t_{\mu\nu} .
\end{equation}
The tensor $\Delta t_{\mu\nu}$ encodes the scale--dependence of the gravitational coupling.
It is given by 
\begin{align} \label{deltatmunu}
\Delta t_{\mu\nu} &= G_k \Bigl(g_{\mu \nu} \square - \nabla_{\mu} \nabla_{\nu}
\Bigl)G_k^{-1},
\end{align}
where $G_k$ and $\Lambda_k$ are now functions of the scale $k$, namely $G_k \equiv G(k)$ and $\Lambda_k \equiv \Lambda(k) $.\\
Note, that the equations (\ref{eq_eomSD1}) with their local covariant structure also arise from 
different approaches and formulations such as
\begin{itemize}
    \item a non-dynamical version of Brans-Dicke theory \cite{Brans:1961sx};
    \item a particular (minimal) case of the broader class of scalar-tensor theories of gravity~\cite{Damour:1992kf,Fujii:2003pa};
    \item 
    $f(R)$ gravity~\cite{Sotiriou:2008rp}, if one imposes that $G(k)$ and $\Lambda(k)$ are functions of the Ricci scalar.
\end{itemize}
Furthermore, we set $\mathcal{L}_m$ to zero in our work and study a universe without matter. Thus $T_{\mu \nu}$ vanishes in the Einsteins equations as well.
At this stage, it is clear that the local nature of the scale-dependence of $G(k)$ plays a crucial role in the equations (\ref{eq_eomSD1}). It is thus clear that any prediction derived from these equations will depend crucially on the ``choice'' of the physical scale $k$.

\subsection{Physically meaningful scale-setting}
\label{subsec_SS}

When we are looking to use $\Gamma_k$ to make a concrete prediction, we need to choose an appropriate expression for the scale $k$ in
terms of the physical parameters of a given problem. These parameters could be e.g. 
position $x^\mu$, time $t$, energy $E$, momentum transfer among others,
\be
k=k(x^\mu,\, E, \, t, \dots).
\ee
This choice is the act of ``scale-setting.'' It is crucial to choose a scale that is physically meaningful for the process we are studying. For instance, in quantum chromodynamics (QCD), when studying processes involving the strong force at a particular energy, we would typically set our scale close to that energy.
Also in the cosmological context, the couplings of a scale-dependent theory, like the gravitational coupling, or the cosmological coupling, become space-time dependent quantities~\cite{Sola:2015wwa,Sola:2017znb,Torres:2017ygl,Ishibashi:2021kmf,Sendra:2018vux,Saueressig:2015xua,Koch:2014cqa,Falls:2012nd,Koch:2013rwa,Bonanno:2001xi,Bonanno:2006eu,Reuter:2006zq}.
For the sake of simplicity we continue our discussion, assuming 
a toy model with completely homogeneous evolution where the
renormalization scale should depend only on the cosmological time \cite{Bonanno:2001xi,Bonanno:2001hi}
\be\label{eq_kt}
k=k(t).
\ee
We are aware that this assumption
can be relaxed by considering 
inhomogeneities, but for this conceptual study already (\ref{eq_kt}) will provide sufficient non-trivial consequences. 
The manner in which this scale is chosen, and the physical reasoning behind it, can influence the results and conclusions one draws from the approach.

It is worth noting that the choice of scale and the associated scale-setting procedure are a source of systematic uncertainty in predictions, and different scale-setting methods might be employed to assess this uncertainty.
Just to name a few:
Variations~\cite{Koch:2010nn,Domazet:2012tw,Koch:2014joa,Koch:2020baj,Koch:2022cta},
 dimensional, symmetry arguments~\cite{Platania:2019kyx,Eichhorn:2021etc,Eichhorn:2021iwq,Held:2021vwd}, energy conditions~\cite{Rincon:2017ayr,Canales:2018tbn,Alvarez:2020xmk,Alvarez:2022mlf,Alvarez:2022wef}.
Moreover, there is an additional theoretical uncertainty that could arise from whether the above scale-setting is to be implemented at the level of~\cite{Reuter:2003ca}
the action~\cite{Koch:2010nn,Domazet:2012tw,Koch:2014joa,Koch:2020baj,Koch:2022cta},
equations of motion~\cite{Bonanno:2020qfu},
or solutions~\cite{Bonanno:2000ep,Bonanno:2006eu,Koch:2013owa,Bonanno:2017zen,Pawlowski:2018swz}. It was in part these systematic uncertainties that lead to criticism~\cite{Donoghue:2019clr}.
Independent of which of the above choices is taken, after the choice the couplings are functions of the physical parameter, e.g. functions of time in the case of (\ref{eq_kt}). As a consequence, the effective equations of motions and their solutions change accordingly.
Before exploring the consequences of (\ref{eq_kt}) and (\ref{eq_eomSD1}) we have to remember that our research question is related to the ``consequence of a fixed point''. Thus, the following two subsections give different notions of fixed points.

\subsection{Scale invariance and fixed points of the dimensionless couplings in the effective action}

Scale invariance (SI) is a property of quantities, systems or phenomena that remain unchanged under a change in scale. 
Depending on the context,
this can refer to spatial scales, temporal scales, or energy scales. 
In the context of quantum gravity and the renormalization group,
one can define SI for different quantities.
Let us consider scale changes of the effective Lagrangian in (\ref{eq_Gammak})
\be\label{eq_Lvar}
\left.k \frac{\partial}{\partial k}{\mathcal{L}}_k(\phi)\right.
=\left(\sum_{i}\beta_i(k){\mathcal{I}}_i(\phi)\right)+
\left(\sum_{l  | j(l)\neq 0}\alpha_l(k){\mathcal{I}}_l(\phi)
k^{j(l)} j(l)\right),
\ee
where the beta functions are defined as
\be
\beta_i
\equiv
k \frac{\partial}{\partial k}\alpha_i.
\ee
A fixed point of the Lagrangian can be defined as
\be\label{eq_fpL}
k \frac{\partial}{\partial k}{\mathcal{L}}_k^*( \phi)=0.
\ee
Since the effective Lagrangian does not have a kinetic term for the scale $k$, relation (\ref{eq_fpL}) is mathematically equivalent to treating $k=k(x)$ as another field and varying the action with respect to this field.
Note that there might also be non-local contributions to the Lagrangian~\cite{Modesto:2017sdr,Modesto:2017hzl,Calmet:2018elv},
since these non-localities make it hard to obtain predictions and since they might also be absent~\cite{Fraaije:2022uhg},
they will not be considered in what follows.
We distinguish two special cases for (\ref{eq_Lvar}):
\begin{itemize}
    \item If the couplings of the Lagrangian are dimensionless ($j(i)=0$) then (\ref{eq_fpL}), 
    combined with the fact that the relation has to hold for arbitrary field content, implies 
\be\label{eq_afp}
\beta_i(\alpha_i=\alpha^*,\dots)
\equiv
\left.k \frac{\partial}{\partial k}\alpha_i\right|_{\alpha_i=\alpha_i^*}= 0.
\ee
This is the traditional definition of SI. It is
    associated to
fixed points $\alpha_i^*$, which
refers to  particular values of the dimensionless coupling constants of a theory for which the beta function vanishes.
\item 
If the  Lagrangian contains dimensionfull couplings  ($j(l)\neq 0$), then (\ref{eq_fpL}), combined with the fact that the relation has to hold for arbitrary field content,  implies for these couplings that
\be
\beta^*_l+\alpha^*_l j(l) k^{j(l)}=0.
\ee
This differs from the traditional definition of a fixed point.
\end{itemize}
Thus, whether a scale-invariant Lagrangian (\ref{eq_fpL}) is equivalent to the traditional notion of a fixed point or not, depends on the dimensionality of the couplings.\\

Following~\cite{Percacci:2007sz},
let us further clarify some related terminology in the realm of SI and coupling fixed points.
There are two kinds of these coupling fixed points that are often discussed: infrared (IR) fixed points and ultraviolet (UV) fixed points. The IR fixed points correspond to long distance or low-energy behavior. If a theory flows to an IR fixed point, it means the theory becomes scale-invariant at large distances or low energies. Physically, phenomena at large scales (like macroscopic scales) would look similar or ``scale'' in a particular way, regardless of how large we go.
The IR fixed points are typically assumed to be unstable such that scale-symmetry can be spontaneously broken, which allows for the existence of massive particles and massless dilatons.
The other type of coupling fixed points is named ultraviolet (UV) fixed points: These correspond to short distance or high-energy behavior. A theory that has a UV fixed point is said to be ``UV complete'' or ``non-perturbatively renormalizable''. This means that at very short distances or high energies, the theory becomes scale-invariant. In the context of quantum gravity, this is of special importance. Einstein's theory of general relativity, which describes gravity at large scales, breaks down at very small scales (like the Planck scale). If a quantum theory of gravity has a UV fixed point, it suggests that the theory is well-defined even at these tiny scales, and there isn't a breakdown of the theory. 
When we speak of a fixed point in the space of theories (often described by a set of parameters or coupling constants), we can imagine perturbing the theory slightly away from this fixed point. The behavior of these perturbations under the RG flow will determine whether they are relevant or irrelevant.
If a perturbation grows as we move to longer length scales (or equivalently, to lower energies) under the RG flow, it is termed ``relevant'', otherwise ``irrelevant''. 

\subsection{Hypothesis: fixed point of the full Lagrangian}

\begin{figure}[hbt!]
\begin{center}
\includegraphics[width=1.0\columnwidth]{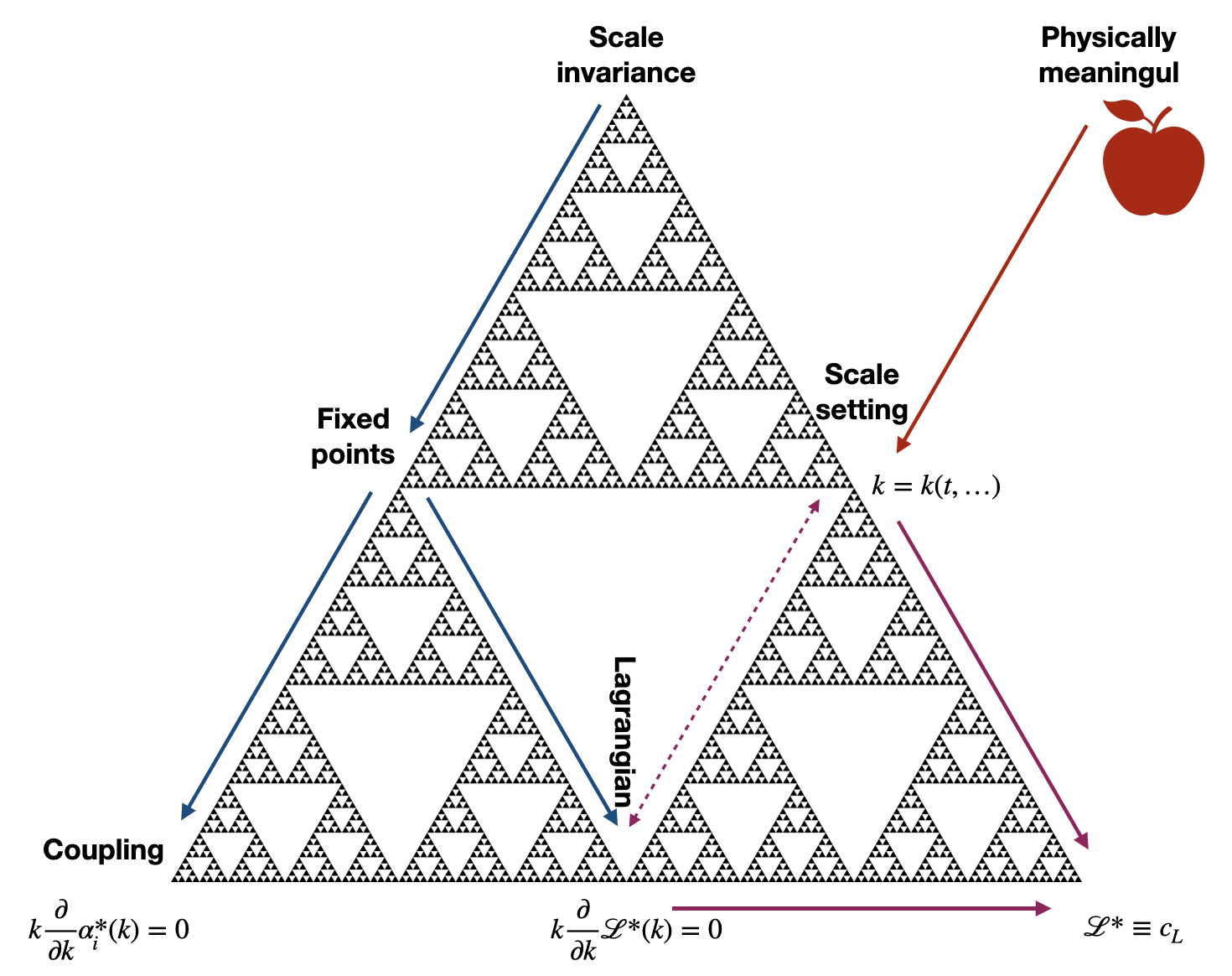} 
\end{center}
\caption{Conceptual map, that shows how SI combined with a physically meaningful prediction motivate ELFP}
\label{fig:Concept}
\end{figure}

The purpose of this study is to combine the two concepts from the preceding subsections. This reasoning is clarified in the conceptual map \ref{fig:Concept}:
\begin{itemize}
    \item On the one hand we have shown that a possible way to study scaling behaviuour and SI
is to do it in terms of the effective Lagrangian and the condition (\ref{eq_fpL}).
\item
On the other hand, we have the ambition to give the effective action a physical meaning. This implies a scale-setting which involves local space-time dependence e.g. (\ref{eq_kt}).
\end{itemize}

Imposing both, a Lagrangian independence of $k$ and a e.g. a space-time dependence of $k=k(t)$, we have to demand that the effective Lagrangian is space-time independent.
Based on this simple observation, let us hypothesize a different class of idealized regime of SI, which we call the effective Lagrangian fixed point (ELFP):

``{\it{The Lagrangian of a theory that has reached  an ELFP will remain constant and thus independent of changes in the scale $k$, and after a scale-setting also independent of continuous physical parameters of the system (such as time for the case of cosmology)}}''
\be\label{eq_cL}
{\mathcal{L}}={\mathcal{L}}^*=c_L.
\ee
Finally, we note that the hypothesis
(\ref{eq_cL}) is specially suitable for a gravitational system, where matter degrees of freedom are negligible. The reason for this is, that the equations of motion for all matter Lagrangian are invariant under a shift ${\mathcal{L}}\rightarrow{\mathcal{L}}+c_L$. This invariance gets lost, when the system is coupled to gravity, a fact which lies in the core of the cosmological constant problem~\cite{Padmanabhan:2006cj}.

\section{The model}\label{Sec_Model}

In order to implement and study the above idea of an ELFP, let us consider a concrete model that is close to classical general relativity. A good starting point for this is an Einstein-Hilbert action with cosmological term.


\subsection{Fixed point condition for the effective Lagrangian}

To explore the physics of
our hypothetical ELFP we  
implement it at the level of the action (\ref{eq_Sk}) in terms of a Lagrange multiplyer $\lambda$
\begin{equation}\label{eq_SkSO}
\left. \Gamma \right|_\mathrm{ELFP}\left(g_{\mu \nu}, k, \lambda\right) \equiv 
 \int \mathrm{d}^4x\sqrt{-g} \left(
 {\mathcal{L}}_k - \lambda  \left( {\mathcal{L}}_k- c_{L} \right)\right).
 \end{equation}  
 This action is manifestly reparametrization invariant. Thus, general covariance of the model is assured.
The action (\ref{eq_SkSO}) can be varied with respect to the three fields $(g_{\mu \nu}, \, k,\, \lambda)$, giving three equations of motion.
First, one can vary with respect to the auxiliary field $\lambda$ giving the ELFP condition
of a constant Lagrangian
\be\label{eq_cL2}
{\mathcal{L}}|_\mathrm{ELFP}= c_L.
\ee
 Second, the equation which arises from variations with respect to the scale-field $k$, is
 \be\label{eq_eomk}
 \frac{\partial  {\mathcal{L}}_k}{\partial k} \cdot (1-\lambda)=0.
 \ee
This equation is trivially solved by (\ref{eq_cL2}).
 The third set of equations arises from variations with respect to the metric field.
 These equations are identical to (\ref{eq_eomSD1}), with the replacement
\be
G_k \rightarrow \tilde G_k\equiv \frac{G_{k}}{1-\lambda}.
\ee
In what follows, we will solve the equations for this effective gravitational coupling and
drop the ``tilde'' notation.


\section{Exploring an ELFP scenario in cosmology} \label{cosmo}

Now, that the stage is set in terms of the action \eqref{eq_SkSO}, we will explore this model in a cosmological context.

\subsection{General considerations}

We assume that:

\begin{itemize}
    \item The universe is in a very early stage, prior to reheating. In this phase, it is commonly assumed that the universe is dominated by inflationary dynamics~\cite{Guth:1980zm}, which means that matter contributions must be negligible. Therefore, we impose $T_{\mu \nu}=0$. We will comment on non-vanishing matter contributions $T_{\mu \nu}\neq 0$ in Subsection \ref{subsec:Beyondanemptyuniverse}.
    \item This regime is a quantum regime, which means that scale-dependence in the gravitational couplings is not negligible, in particular: $G=G_k$ and $\Lambda = \Lambda_k$. Due to limited algebraic and computational power, we restrict to these two couplings in the gravitational sector.
    \item The universe is in a extremely symmetric state, which means that the metric is homogeneous and only time dependent, just like the SD couplings
    \bea
    g_{\mu \nu}&=&g_{\mu \nu}(t),\\
    G_k&=&G(t),\\
    \Lambda_k&=& \Lambda(t).
    \eea
    Further, and this is crucial, the scale setting is maximally insensitive to the choice of the quantum scale, which is implemented by (\ref{eq_cL2}).
    \item By working with the Einstein-Hilbert truncation (\ref{eq_Gammak}) and the corresponding equations (\ref{eq_eomSD1}) we implicitly make the simplifying assumption that higher curvature contributions can be cast into a pure scale-dependence effect similar to the 
    dynamical equivalence of $f(R)$ gravity in Jordan and Einstein frames~\cite{Chakraborty:2018ost}.
\end{itemize}

There is a remarkable consequence of these conditions and the ELFP relation~(\ref{eq_cL2}). It
allows us to solve the equations directly for $G=G(t)$ and $\Lambda=\Lambda(t)$, as will be shown in the next section.

\subsection{Ansatz}

The goal of this work is to provide a proof-of-principle, by exploring a theoretical toy model and using simplifying assumptions to make the model manageable. Two such assumption are the conditions of homogeneity and flatness ($\kappa =0$), which are implemented through the following ansatz for the line element
\be\label{eq_ds2}
ds^2= - dt^2+ a^2(t) (dx^2+dy^2+dz^2).
\ee
For this line element
 the equations of motion (\ref{eq_eomSD1}) simplify to the SD Friedmann
 equations 
\bea\label{eq_FriedSD1}
\Lambda&=&-\frac{3 \dot a(a \dot G - G \dot a)}{a^2 G},\\ \label{eq_FriedSD2}
\Lambda-\frac{\dot a^2}{a^2}&=&\frac{2 \dot a \dot G-2 G \ddot a}{G a}+
\frac{G \ddot G - 2 \dot G^2}{G^2}
\eea
and
the ELFP condition
\be\label{eq_ELFP2}
\frac{6(\dot a^2+ a \ddot a)}{a^2 G}- 2\frac{\Lambda(t)}{G(t)}=c_L.
\ee
One verifies that (\ref{eq_FriedSD1}) and (\ref{eq_FriedSD2}) simplify to the usual ``classical'' Friedman equation
\be\label{eq_FriedCl}
\frac{\dot a^2}{a^2}=\frac{\Lambda_0}{3},
\ee
when the couplings become constants
$G \rightarrow G_0$ and $\Lambda \rightarrow \Lambda_0$, which is a solution of (\ref{eq_FriedSD1}) -(\ref{eq_ELFP2}) .
The ``classical'' equations (\ref{eq_FriedCl})
are solved by an exponential growth of the scale-factor
\be \label{aclassic}
a_{cl}(t)= a_0 \exp\left(t \sqrt{\frac{\Lambda_0}{3}}\right)\,.
\ee


The relations (\ref{eq_FriedSD1}), (\ref{eq_FriedSD2}) and (\ref{eq_ELFP2}) correspond to a higher-order system, but, as it will be shown, the system can be integrated analytically.

At this point, several key observations need to be highlighted. First, let us emphasize the significance of the additional terms in the modified Friedman equations, such as those in \eqref{eq_FriedSD1}, \eqref{eq_FriedSD2}, and \eqref{eq_ELFP2}. It is important to note that all additional terms proportional to $\dot{G}(t)$ and $\ddot{G}(t)$ arise from scale-dependent gravity, meaning that Newton's coupling—as well as the cosmological coupling—can now vary. Consequently, these new terms are not directly related to any spatial anisotropy or inhomogeneity of the universe, although such a correlation might arise indirectly.
As mentioned earlier, if gravity operates with constant couplings, then the extra terms should not appear. However, if gravity does not function uniformly across our universe, particularly in an empty universe, it is important to clarify that this does not necessarily lead to anisotropy or inhomogeneity of the universe, nor does it imply any violation of the cosmological principle.

\subsection{Solving the equations} \label{sec_solvingEq}

For solving, the equations (\ref{eq_FriedSD1}, \ref{eq_FriedSD2}, \ref{eq_ELFP2}), one can start with
algebraic identities for $\Lambda$.
For example, 
Eq.~(\ref{eq_FriedSD1}) can be read as identity for $\Lambda$.
Alternatively one can also read $\Lambda$ from Eq.~(\ref{eq_ELFP2}) as
\be\label{eq_tmp1}
\Lambda(t)=\frac{6(\dot a^2+ a \ddot a)}{2 a^2 }-G\frac{c_L}{2}.
\ee
Even though, the two conditions 
(\ref{eq_FriedSD1}) and (\ref{eq_tmp1})
do not look identical, their difference must vanish. This difference can be 
translated into a condition on $\dot G$
\be\label{eq_tmp2}
\dot G = G \frac{c_L a G- 6 \ddot a}{6 \dot a}.
\ee
Deriving the condition (\ref{eq_tmp2}) with respect to $t$, gives an analogous
condition for $\ddot G$.
Now, one use this on the two conditions and (\ref{eq_tmp1}) on (\ref{eq_FriedSD1}), to get a decoupled differential equation
\be\label{eq_tmp3}
\frac{G^2}{\dot a} \left( 
2 \dot a^3 - a \dot a \ddot a - a^2 \frac{d^3a}{dt^3} 
\right)=0.
\ee
We can recognize the time derivative of the Ricci scalar in the term between parenthesis,
\be
\label{eq_Ricci}
\frac{R(t)}{6}= \frac{\dot a^2}{a^2}+ \frac{\ddot a}{a},
\ee
therefore an important class of  solutions of (\ref{eq_tmp3}), described by $\dot R =0$, corresponds to
\be\label{eq_Jcd}
R = c_R,
\ee
where $c_R$ is an integration constant.

We will present two alternative complementary ways to express the new solutions. Albeit it could be self-evident, let us reinforce that a concrete form of the scale factor plays a critical role at the moment of identifying the corresponding integration constants. In what follows, we will consider a solution written in terms of exponential functions, trying to mimic the classical scale factor representation. Thus, by replacing Eq. \eqref{eq_Jcd} we can find a solution of Eq. \eqref{eq_Ricci}:
\begin{align} \label{at2}
a(t) &= \frac{1}{\sqrt{2}}c_{2} e^{\frac{t}{2}\sqrt{\frac{c_{R}}{3}}}
\sqrt{
e^{  c_{1} \sqrt{ \frac{ c_{R} }{ 3 } } }
+
e^{ -  (2t + c_{1}) \sqrt{ \frac{ c_{R} }{ 3 } } },
}
\end{align}
and, subsequently, we can replace Eq. \eqref{at2} into Eq. \eqref{eq_tmp2} to get a compact expression for the Newton's coupling, which is
\small
\begin{align} \label{eq_Gt2}
    G(t) &= \frac{c_{R} \sqrt{Y(t)(Y(t)^2 + 1)} }{
c_{L} \sqrt{Y(t)} 
    \Bigl(
    \sqrt{Y(t)^2 + 1} - (Y(t)^2 - 1) {}_{2}F_{1} \Bigl( \frac{1}{4}, \frac{1}{2}; \frac{5}{4}; -Y(t)^2  \Bigl)
    \Bigl)
    +
    c_{3} c_{R} (Y(t)^2 - 1)
    },
\end{align}
\normalsize
where we have defined the auxiliary function $Y(t)$,
\begin{align}
    Y(t) \equiv \exp \left(\sqrt{\frac{c_{R}}{3}}(t + c_{1}) \right),
\end{align}
and finally, we use Eq.\eqref{eq_tmp1},
to obtain the cosmological coupling 
\begin{align}
    \Lambda(t) &= \frac{1}{2} \Bigl( c_{R} - c_{L} G(t) \Bigl), \label{eq:lambdasol}
\end{align}
where the gravitational coupling is given in (\ref{eq_Gt2}).
%
At this point it is clear that a trivial identification of the integration constants is a difficult task, at least in the present representation. However, if we consider the scale factor in terms of hyperbolic functions, we can clearly identify Newton's constant (at $t=t_i$).
Inserting (\ref{eq_Ricci}) in (\ref{eq_Jcd}) and integrating it  yields the solution for the scale factor
\be\label{eq_atsol0}
a(t)= c_2  \Bigg(  \cosh{ \left[\sqrt{\frac{c_R}{3}}\Bigl(c_1 + t \Bigl)\right] } \Bigg)^{1/2} ,
\ee
where $c_{1,2}$ are the corresponding integration constants.
This solution, Eq. (\ref{eq_atsol0}), can be inserted back
into (\ref{eq_tmp2}) leading to an ordinary differential equation for $G(t)$,
\begin{equation}\label{eq_dotG}
 \dot{G}(t) = F_1(t) G(t) + F_2(t) G(t)^2\,,
\end{equation}
where
\begin{align}\label{eq_F1}
 F_1(t) =& -\sqrt{c_R/3} \mathrm{coth} (\sqrt{c_R/3}(t+c_1)) + \frac{\sqrt{c_R/3}}{2} \mathrm{tanh} (\sqrt{c_R/3}(t+c_1))\,,\\
 F_2(t) =& \frac{c_L}{3\sqrt{c_R/3}} \mathrm{coth} (\sqrt{c_R/3}(t+c_1)) \,.\label{eq_F2}
\end{align}

The equation has the Bernoulli form and therefore the general solution can be written as
\begin{equation}
\frac{1}{G(t)} = \exp  \left(-\int_{t_i}^t dt' F_1 \right) \left[\frac{1}{G_i} - \int_{t_i}^t dt' F_2 \exp  \left(\int_{t_i}^{t'} dt'' F_1 \right) \right]\,, \label{1overG}
\end{equation}
where $G_i$ is an integration constant that has the meaning of the value of $G(t)$ at the time $t_i$, $G(t_i) = G_i$. The following expressions are useful:
\begin{equation}
 \exp  \left(\int_{t_i}^t dt' F_1 \right) =\frac{\mathcal{F}(t;c_R,c_1)}{\mathcal{F}(t_i;c_R,c_1)}\,, \label{expF1}
\end{equation}
where
\begin{equation}
 \mathcal{F}(t;c_R,c_1) = \frac{\sqrt{\mathrm{cosh} (\sqrt{c_R/3}(t+c_1))}}{\mathrm{sinh} (\sqrt{c_R/3}(t+c_1))}\,,
\end{equation}

and
\begin{align}
 \int_{t_i}^t dt' F_2 \exp  \left(\int_{t_i}^{t'} dt'' F_1 \right) =& \frac{c_L}{|c_R|}\left[ 1 - \frac{\mathcal{F}(t;c_R,c_1)}{\mathcal{F}(t_i;c_R,c_1)}\right. \nonumber\\
 -& \left. \frac{i}{\mathcal{F}(t_i;c_R,c_1)} (E^{(F)}(i\sqrt{c_R/12}(t+c_1),\ 2) \right.\nonumber\\
 -& \left. E^{(F)}(i\sqrt{c_R/12}(t_i+c_1),\ 2) \right]\,, 
 \label{expF2}
\end{align}
where
\begin{equation}
 E^{(F)}(x,m)=\int_0^x d\theta \frac{1}{\sqrt{1-m \sin ^2 \theta}}\,,
\end{equation}
is an elliptic function of the first kind. Using (\ref{expF1}) and (\ref{expF2}) in (\ref{1overG}) we get
\begin{align}
 \frac{1}{G(t)} =& \frac{\mathcal{F}(t_i;c_R,c_1)}{\mathcal{F}(t;c_R,c_1)} \left[ \frac{1}{G_i} - \frac{c_L}{|c_R|} +\frac{c_L}{|c_R|} \frac{\mathcal{F}(t;c_R,c_1)}{\mathcal{F}(t_i;c_R,c_1)}  \right.\nonumber\\
 +& \left. \frac{i c_L}{|c_R|\mathcal{F}(t_i;c_R,c_1)}  (E^{(F)}(i\sqrt{c_R/12}(t+c_1),\ 2) \right.\nonumber\\
 -& \left. E^{(F)}(i\sqrt{c_R/12}(t_i+c_1),\ 2)) \right]\,.\label{1overGclosed}
\end{align}

We will see that the behavior of $G(t)$ depends crucially of the values of $G_i$, $c_L$ and $c_R$. 

\subsection{Physical interpretation of the integration constants} \label{section_redefiningConstants}

The above solution has
five free parameters which consist of one actual parameter of the Lagrangian $c_L$ and four integration constants $c_R$, $c_1$, $c_2$ and $G_i$.
In this subsection
these parameters will be brought into a more intuitive form.

Let's first study the solution $a(t)$. 
It is straight forward to see that (\ref{eq_atsol0}) describes a bouncing scale factor and that $c_1$ defines
 the time of the bounce, thus we write
\be\label{eq_id1}
c_1=-t_b.
\ee
The condition
\be
\lim_{t_b\rightarrow -\infty} a(t) =  a_{cl}(t)
\ee
allows us to 
rewrite the two constants $c_R$ and $c_2$.
From the Ricci scalar of the classical solution
\be
R|_{a_{cl}}=4 \Lambda_0,
\ee
we relabel the corresponding integration constant
\be
c_R \equiv 4 \Lambda_0.
\ee
Next,
fix the constant
\be\label{eq_id2}
c_2= \sqrt{2} a_0  \exp{\left(\sqrt{\frac{\Lambda_0}{3}} t_b \right)}.
\ee
After these re-definitions the scale-factor reads in terms of the integration constants
$\Lambda_0, t_b$ and $a_0$ 
\be
a=\sqrt{2}a_0 \exp{\left( \sqrt{\frac{\Lambda_0}{3}} t_b \right)}
\left(\cosh{2\sqrt{\frac{\Lambda_0}{3}}(t - t_b)}\right)^{1/2}.
\ee
Similarly we can proceed with the gravitational coupling.
For this, we keep in mind that there are four time-scales involved in the definition of $G(t)$: $t_b < t_i<t$.

For large $t$ the gravitational coupling drops to zero, $\lim_{t\rightarrow \infty} G(t)=0$, see figures~(\ref{fig:gridGt_1},\ref{fig:gridGt_2}). However, for a short period of time, $G(t)$ has a plateau behavior. The time length of the plateau depends on the parameters, as we see it in the figure. There is an special case in which the time length of the plateau get enhanced and it occurs when $c_L = 2 \Lambda_0/ G_i$.
To get a better understanding of this behavior,  we explore the regime $t_b\ll t_i,t$, where we can approximate the functions (\ref{eq_F1}, \ref{eq_F1})
\bea
F_1|_{t_b\ll t_i}&=&-\sqrt{\frac{\Lambda_0}{3}}\\
F_2|_{t_b\ll t_i}&=& \frac{c_L}{2 \sqrt{3 \Lambda_0}}.
\eea
In this regime
the differential equation (\ref{eq_dotG}) reads
\be
\frac{d}{dt}\left( \frac{1}{G}\right)= - \sqrt{\frac{\Lambda_0}{3}} \left( \frac{1}{G}\right) -\frac{c_L}{2 \sqrt{3 \Lambda_0}}.
\ee
We can solve this linear differential equation with the initial condition $G(t_i)=G_i$, giving
\be
\left( \frac{1}{G}\right)= \exp \left( \Delta t\sqrt{\frac{\Lambda_0}{3}}\right)
\left( \frac{1}{G_i}-\frac{c_L}{2 \Lambda_0}\right)
+ \frac{c_L}{2 \Lambda_0},
\ee
where $\Delta t\equiv t-t_i\gg t_b$.
There are two main cases, firstly if $c_L\neq 2 \Lambda_0/ G_i$,  then:
\be
\lim_{\Delta t\rightarrow 0} G(t) = G_i\,,  \quad \lim_{\Delta t\rightarrow 0} \Lambda(t) = 2\Lambda_0-\frac{c_L G_i}{2}\\
\ee
and the solution has the following asymptotic behavior
\be
\lim_{\Delta t\rightarrow \infty} G(t) = 0\,, \quad \lim_{\Delta t\rightarrow \infty} \Lambda(t) = 2\Lambda_0\,.\label{eq_GLasym}\\
\ee
Also, if $G_i < 2 \Lambda_0/c_L$, then $G(t)$ continuously but quickly decrease to a vanishing value (figure \ref{fig:gridGt_1}). If $G_i > 2 \Lambda_0/c_L$, then $G(t)$ may blow up very quickly to then emerge from $-\infty$ and approach a vanishing value very quickly (figure \ref{fig:gridGt_2})
Secondly, in the fine-tuned case, $c_L = 2 \Lambda_0/ G_i$, we have a regime where both couplings appear to be approximately constant
\be\label{eq_GLbranch}
 G(t) = \frac{2 \Lambda_0}{c_L} \,, \quad \Lambda(t) = \Lambda_0\,.
 \ee
This case is special because the time duration of the plateau of $G(t)$ gets enhanced. We denote the time length at which $G(t)$ is approximately constant by $\Delta t_d$. We found it being proportional to
    \begin{equation}
         \Delta t_d\sim \frac{1}{\sqrt{c_L}}= 
        \sqrt{\frac{G_i}{2 \Lambda_0}}\,.
    \end{equation}
We determined by a numerical analysis by looking at the time at which $G(t) = (1/2) G_i$. In figure~\ref{fig:Gt_timescale}, we vary $\Lambda_0$ and we can observe that a smaller $\Lambda_0$ implies a later drop of the gravitational coupling.

To summarize this section, an intuitive parametrization of our four integration constants can be given in terms of 
the scale factor normalization at late times $a_0$, the expansion factor
at late times $\Lambda_0$, the bouncing time $t_b$, and the value of the gravitational coupling at an intermediate time $G(t_i)=G_i$.
For this scenario the couplings approach the values given in (\ref{eq_GLasym}).
If further, $G_i= \frac{2 \Lambda_0}{c_L}$ and we take the singular limit $t_i\rightarrow \infty$ with $t_i\gg t_b$, the couplings approach finite constant values (\ref{eq_GLbranch}).

\subsection{Curvature invariants}

One can further study 
the curvature invariants of our solution.
The Ricci scalar takes the same value as in the classical case, $R= 4 \Lambda_0$, however higher curvature scalars like
the Kretschmann scalar are sensitive to the deviations from the classical case:
\be \label{kretsch}
    R^{\mu \nu \alpha \beta} R_{\mu \nu \alpha \beta}=
    \frac{8}{3}\Lambda_0^2 \left( 1 + \text{sech} \left(2\sqrt{\frac{\Lambda_0}{3}} (t-t_b) \right)\right).
\ee
\begin{figure}[ht!]
    \begin{center}
\includegraphics[width=0.71\columnwidth]{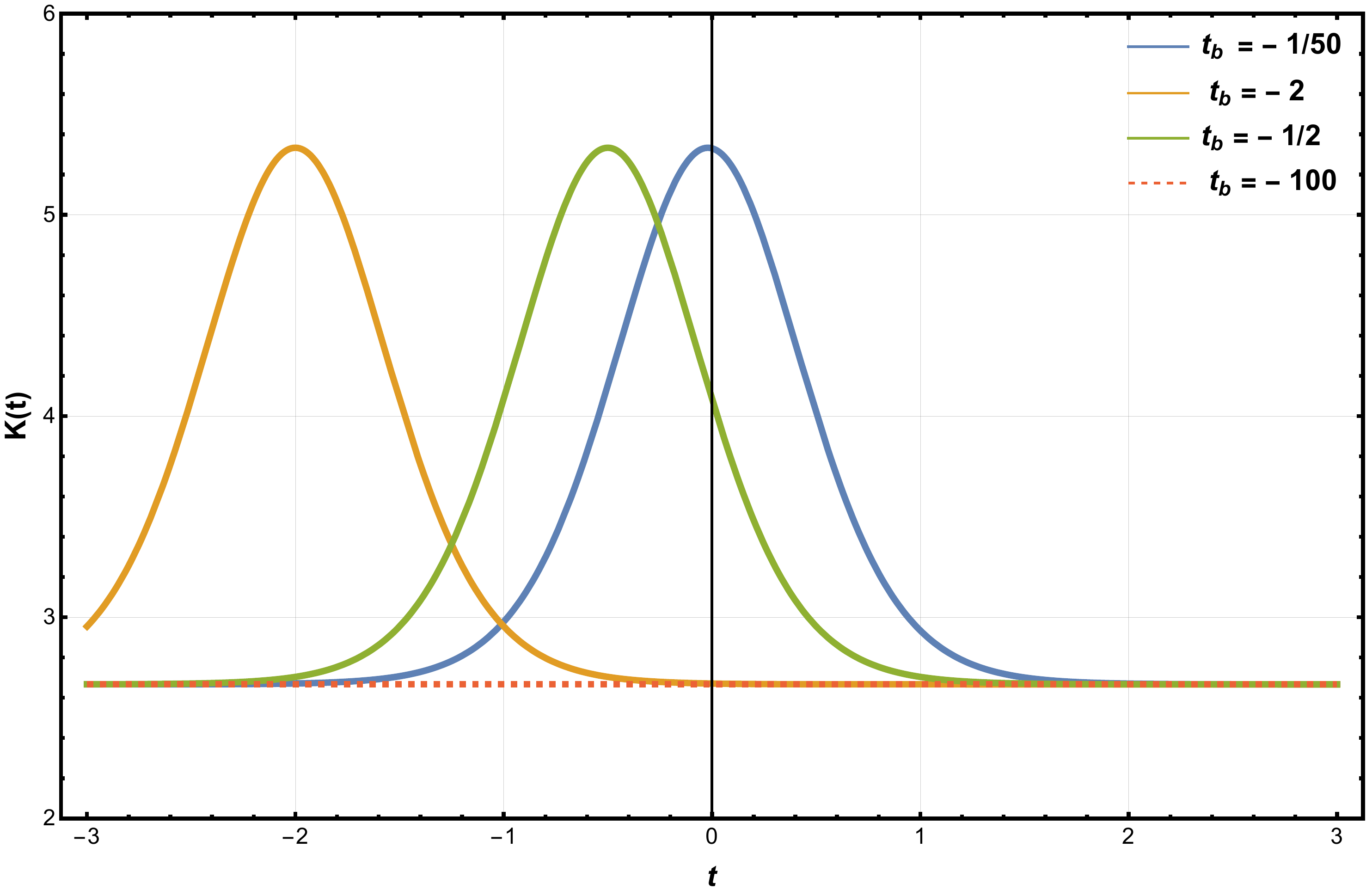} 
    \end{center}
    \caption{Visualization of the Kretschmann scalar in Eq.~(\ref{kretsch}) for four different choices of the bouncing time $t_b$. In all the scenarios, $\Lambda_0$ is set to one.}
    \label{fig:Kretsch}
\end{figure}
In figure~\ref{fig:Kretsch}, we observe two interesting features. First, one immediately notices that as the bouncing time is pushed back in time, the contribution from the second term in Eq.~(\ref{kretsch}) becomes negligible for late times and the solution converges to the classical case as we have highlighted this in the figure with dashed line. Second, when time $t$ reaches the bouncing time $t_b$, the argument in the hyperbolic function becomes zero and we have a maximum and again, as $t$ moves forward or backwards away from $t_b$, we recover the classical Kretschmann scalar. In terms of the statefinder parameters the model is described by a straight line in the $(q,r)$-plane with slope unity that finishes in the classical point $(q,r)_\mathrm{cl} = (1,1)$
\section{Discussion}
\label{sec_Discussion}

In this final section, we  revisit and discuss the preceding results.

\subsection{The dynamical variables}

Our results provide several interesting features for the functions $a(t),\, G(t),\,  \Lambda(t)$. Since analytical structure of  $G(t)$ is more involved than the structure of the other functions, the discussion of $G(t)$ will require somewhat more space than the discussion of the other two functions.

\begin{itemize}
\item $a(t)$: In the scale factor we notice truly remarkable features that can be read from the figure \ref{fig:at}. We realize at large times the function $a(t)$ shows the expected classical exponential evolution. However, when evolving backwards in time, the curve flattens, reaches a minimum and then grows again. We observe that, for sufficiently large times the solution becomes harder to distinguish from a purely classical one.
\begin{figure}[ht!]
    \begin{center}
    \includegraphics[width=0.71\columnwidth]{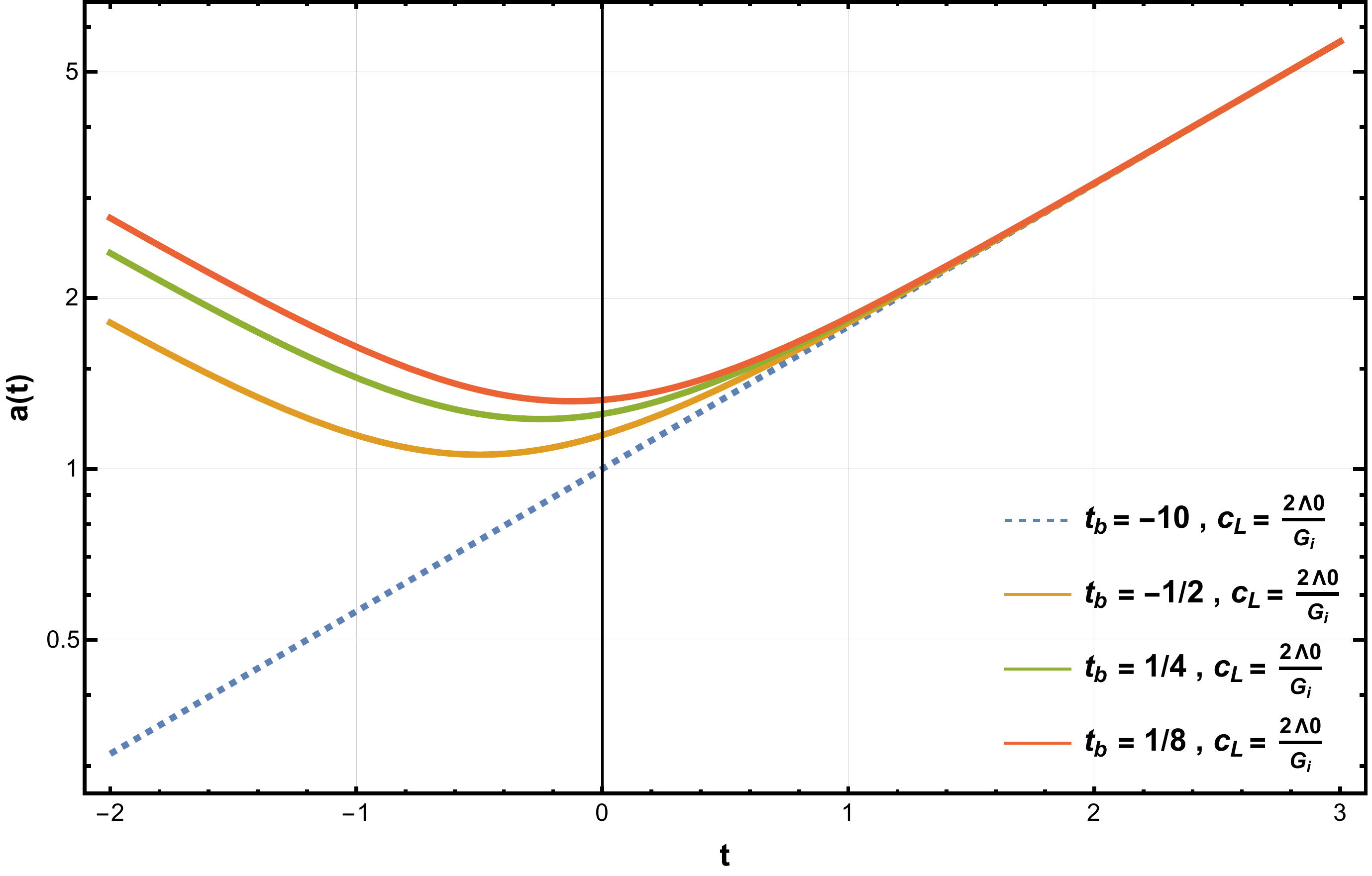} 
    \end{center}
    \caption{Visualization of the evolution of the scale factor for four different scenarios all with $c_L = 2 \Lambda_0 / G_i$. The y-axis is in logarithmic scale. Both $\Lambda_0$ and $G_i$ are set to one in all the four scenarios.}
    \label{fig:at}
\end{figure}

\item $G(t)$:
The time dependence of the gravitational coupling is shown in figures (\ref{fig:gridGt_1},\ref{fig:gridGt_2}). In the figure, we have defined three different categories by keeping the ratio $2\Lambda_0 /G_i$ and defining $c_L$ to be equal, greater and smaller than $2\Lambda_0 /G_i$. We notice some features from the behaviour of the coupling. First, the coupling drops to zero at very large times and this behaviour is unavoidable. However, as mentioned in section~\ref{sec_solvingEq}, by fine-tuning the solution of $G(t)$, it is possible to keep the coupling constant for some time before it drops to zero. This can be seen from the scenarios in the middle column of the figure. We also notice that in all scenarios
the coupling diverges as the time $t$ approaches the bouncing time $t_b$. 
The reason for this is that the function $\mathcal{F}(t,c_R,t_b) = \frac{\sqrt{\text{cosh} ( \sqrt{cR/3} (t-tb)}}{\text{sinh} ( \sqrt{cR/3} (t-tb)}$ diverges, causing the coupling $G(t)$ to diverge as well. \\
Also, as discussed in section~\ref{section_redefiningConstants}, for late times and when $t_b \ll t,t_i$, one can use approximations 
to understand why the couplings become approximately constant. However, one needs to be careful with this approximation. In order to use this asymptotic feature of the model, the approximation has to be made at the level of the differential equations as discussed. Once the full differential equations, without the approximations in section~\ref{section_redefiningConstants},  are solved and the solution (Eq.~\ref{1overGclosed}) is found, the coupling necessarily drops to zero at some late enough time.\\

\begin{figure}[H]
    \begin{center}
    \includegraphics[width=1\columnwidth]{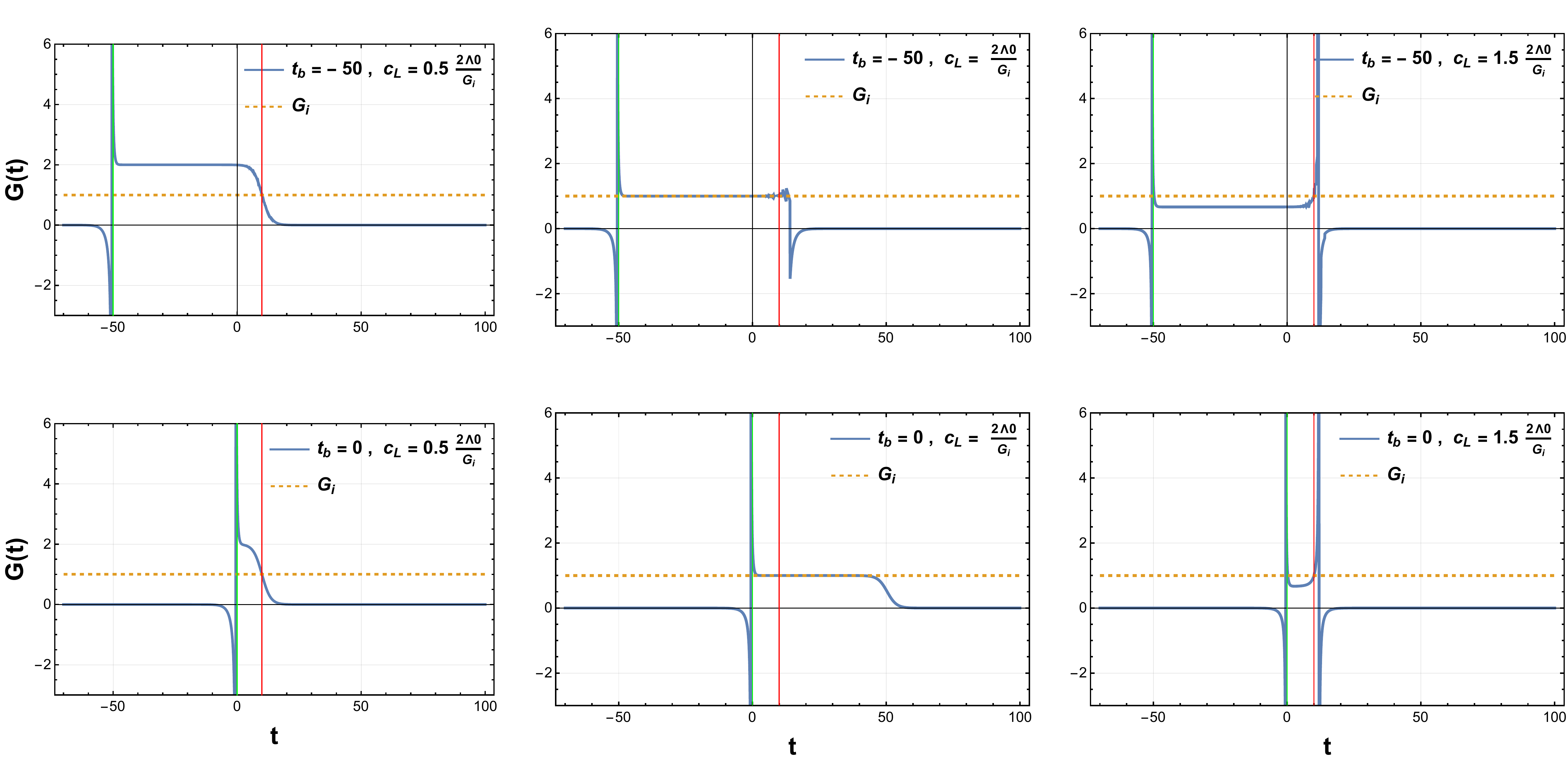} 
    \end{center}
    \caption{Visualization of the evolution of the scale-dependent gravitational coupling constant for different scenarios with  $t_b < t_i$ . In the matrix, the rows correspond to the variation of the bouncing time $t_b$ and the columns correspond to the three different cases $c_L < 2 \Lambda_0 / G_i$, $c_L = 2 \Lambda_0 / G_i$ and $c_L > 2 \Lambda_0 / G_i$, respectively. The vertical green line indicates the chosen bouncing time and the vertical red line indicates the chosen initial time where the value of $G_i$ is fixed. In all the figures, $G_i$ and $\Lambda_0$ are set to one.}
    \label{fig:gridGt_1}
\end{figure}

\begin{figure}[H]
    \begin{center}
    \includegraphics[width=1\columnwidth]{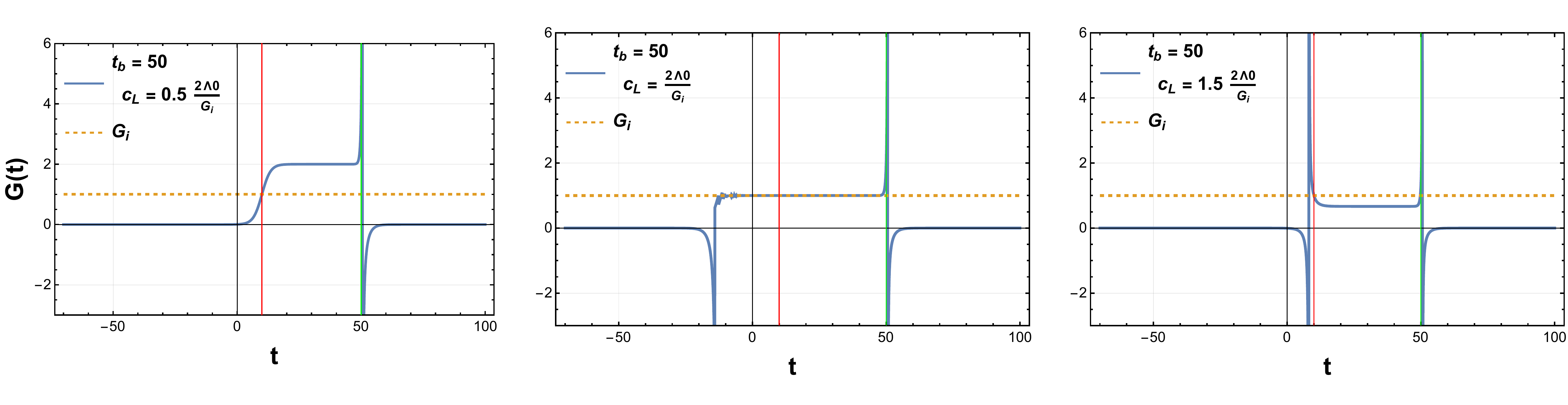} 
    \end{center}
    \caption{Same as Figure \ref{fig:gridGt_1} but with $t_b > t_i$.}
    \label{fig:gridGt_2}
\end{figure}

\begin{figure}[H]
    \centering
    \includegraphics[width=1\columnwidth]{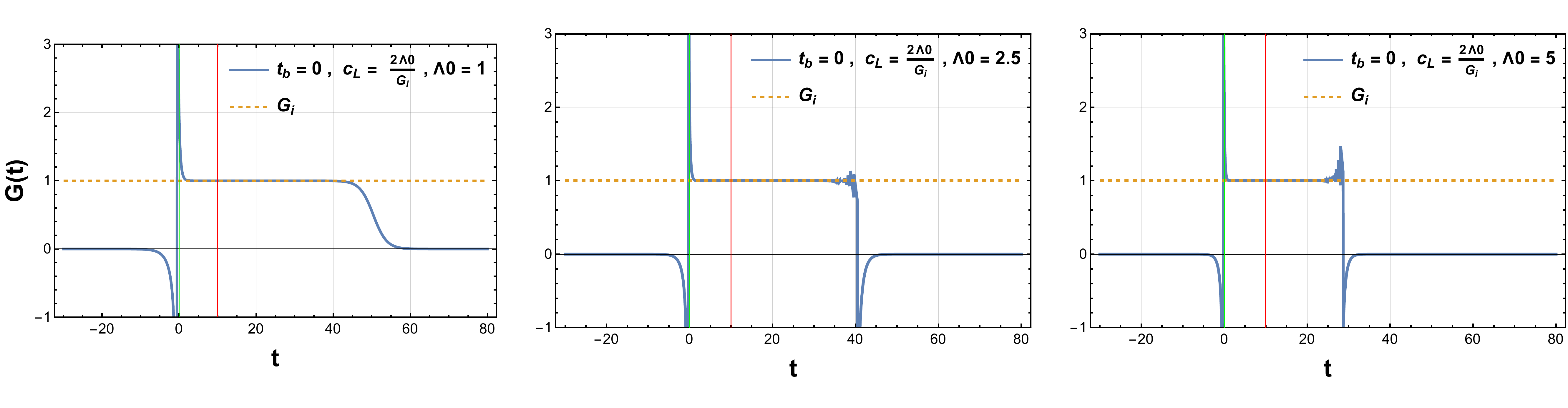}
    \caption{Visualization of $G(t)$ in the case where $\Lambda_0$ is being varied, illustrating the fact that a smaller $\Lambda_0$ makes $G(t)$ non-zero for a larger period of time. $G_i$ and $t_i$ are set to 1 and 10, respectively.}
    \label{fig:Gt_timescale}
\end{figure}
One notes further that for $c_L=0$ the analytical form of $G(t)$ gets particularly simple, but apart from this, the qualitative behaviour remains the same.


\item $\Lambda(t)$:
The evolution of the cosmological term is shown in figure~\ref{fig:Lt}. In the evolution backwards and also forwards in time one notes that the cosmological term reaches a constant  value of $2\Lambda_0$ due to the behaviour our of the scale dependent coupling $G(t)$. Also, due to the behaviour of $G(t)$, when $G(t)$ diverges, so does $\Lambda (t)$ but with a different sign because of the minus sign in Eq.~(\ref{eq:lambdasol}).
%
\begin{figure}[H]
    \begin{center}
    \includegraphics[width=0.7\columnwidth]{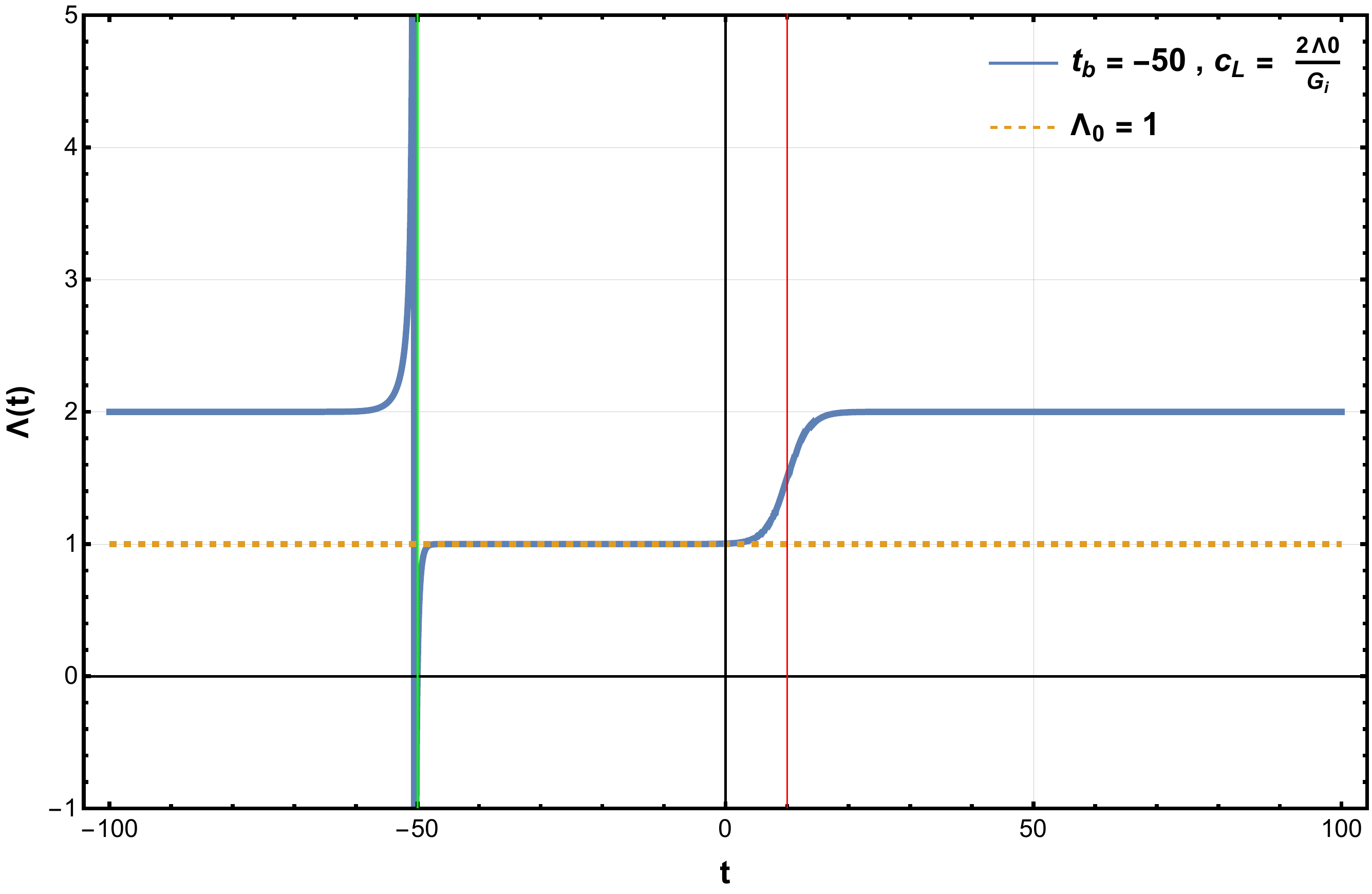} 
    \end{center}
    \caption{Visualization of the evolution of the scale-dependent cosmological constant for a scenario with $c_L = 2 \Lambda_0 / G_i$. The vertical green line corresponds to the bouncing time and the vertical red line corresponds to the initial time where the value of $G_i$ is being fixed. The values of $G_i$ and $\Lambda_0$ are set to one.}
    \label{fig:Lt}
\end{figure}
%

\end{itemize}

\subsection{Towards a dynamical systems perspective}

Attractors are fundamental concepts in the study of dynamical systems. An attractor is a set or a point towards which a dynamical system evolves over time, regardless of the starting conditions (within a given region). Attractors represent the long-term behavior of systems. There are various types of attractors, and they can be distinguished based on their structure and the nature of the dynamical system:
Fixed points, attractors, limit cycles, limit torus, strange attractors.

It is interesting to analyze the above results from the perspective of such a dynamical system. The dynamical variables are $a(t)$ and $G(t)$. They span the four dimensional phase space ${\bf{X}}\equiv\left\{a,\, \dot a, G,\, \dot G\right\}(t)$.
Analyzing the time evolution 
of our solution we find that 
\be
{\bf{X}}_{FP1}=
\left\{a_{cl},\, \dot a_{cl}, 0,\, 0\right\}(t\rightarrow \infty)
\ee
is a stable fixed point (attractor) of the dynamical system.  
There further seems to be another, unstable fixed point
\be
{\bf{X}}_{FP2}=
\left\{a_{cl},\, \dot a_{cl}, \frac{2 \Lambda_0}{c_L},\, 0\right\}(t\gg t_b).
\ee
Here, it is important to note that a priory fixed points of RG systems are not related to fixed points of dynamical systems, because the latter are understood as characteristics of the time evolution of a system like $\left\{a,\, \dot a, G,\, \dot G\right\}(t)$, while the former are related to scale transformations of a different system $\Gamma(k,\dots)$.
This clear separation of the two concepts gets however bridged as soon as one seeks a physical interpretation of the RG system in terms of a scale setting (\ref{eq_kt}) as discussed in section~\ref{subsec_SS}.
It is highly intriguing to observe that demanding an ELFP for an effective action
leads to a dynamical system which has one attractive and one repulsive fixed point.

It will be an interesting subject of future studies, to explore this apparent scale-setting induced link between RG fixed points and the attractors of the corresponding dynamical systems. In particular since such attractors have also been found in the broader context of certain tensor-scalar theories~\cite{Damour:1992kf}.

\subsection{Parallels to bouncing cosmology models}
\label{sec:Parallelstobouncingcosmologymodels}
Bouncing cosmologies refer to models of the universe in which there was a contraction phase that preceded the current expansion phase, and instead of a singularity (as in the Big Bang model), there was a ``bounce'' at some minimum scale. Essentially, these models propose that our universe underwent a contraction, reached a small but finite size, and then began expanding, which is the phase we observe today.
They are interesting because they can serve for avoiding singularities, provide an
alternative to inflation,
shed light on
quantum gravity, provide
cyclic universes, and at timed provide observational clues.

The results presented above show intriguing similarities with completely different approaches to gravity which also produce such a bouncing cosmology. These are for example, Horava-Lifshitz gravity  ~\cite{Horava:2009uw,Horava:2009if},
$f(R)$ gravity~\cite{10.1093/mnras/150.1.1,Nojiri:2009kx}, $f({\mathcal{G}})$ gravity~\cite{Astashenok:2015haa}, $f(T)$ gravity~\cite{Geng:2011aj,Maluf:2013gaa}
and even loop quantum cosmology~\cite{Finelli:2001sr,Laguna:2006wr,Corichi:2007am,Bojowald:2008pu}
(for a summary see \cite{Nojiri:2017ncd}).
Since the ELFP was proposed from a perspective of scale invariance, it is interesting that the bounce we find seems to reflect the essence of another cosmological model, whose construction principle is conformal symmetry~\cite{Gurzadyan:2013cna}. It is interesting to make a detailed comparison between $f(R)$ models with bounce and our ELFT proposal, which could be further explored in future studies.

\subsection{The role of matter}
\label{subsec:Beyondanemptyuniverse}

In our analysis, we have set $\kappa=0$ and $T_{\mu \nu} = 0$. Thus, our results correspond to a universe without radiation, matter or curvature term. At later stages, posteriour to reheating, the presence of matter became the driving force the the dynamical evolution of our universe, starting with radiation.
Given an energy density $\Omega_r$ associated to radiation, 
leads to an additional term in eqs.~(\ref{eq_FriedSD1},\ref{eq_FriedSD2},\ref{eq_ELFP2}). 
We get
\begin{equation}
    \label{eq_FriedSD1_RD}
    \Lambda=-\frac{3 \dot a(a \dot G - G \dot a)}{a^2 G} - 8\pi \rho_0 G/a^4
\end{equation}
\begin{equation}
    \label{eq_FriedSD2_RD}
    \Lambda-\frac{\dot a^2}{a^2}=\frac{2 \dot a \dot G-2 G \ddot a}{G a}+ \frac{G \ddot G - 2 \dot G^2}{G^2} + 8\pi \rho_0 G/a^4
\end{equation}
\begin{equation}
    \label{eq_ELFP2_RD}
    \frac{6(\dot a^2+ a \ddot a)}{a^2 G}- 2\frac{\Lambda(t)}{G(t)}+\mathcal{L}_m =c_L.
\end{equation}
In the above equations $\rho_0$ is defined as $\rho_0=\frac{3 H_0^2 \Omega_r}{8 \pi G_N}$ with $\Omega_r$ being the density parameter of radiation, $G_N$ the newton coupling and $H_0$ the Hubble's constant.In the simplest case, the Lagrangian $\mathcal{L}_m$ contains only the photon fields $F_{\mu \nu}F^{\mu \nu}$ but one can also include other fundamental fields with respect to the era of radiation domination of the universe. However, this analysis presents challenges that are beyond the scope of this work. The differential equations of section~\ref{sec_solvingEq}, which are used to obtain the couplings as well as the scale factor, become much more complicated. For instance, on the r.h.s of eq.~(\ref{eq_tmp3}) we will have $\frac{8 \pi \rho_0}{3} G(t)^3 \left( 1 - 1/a(t)^2 \right)$. Thus, on cannot simply use $R = c_R$ and the definition of Ricci scalar to find the scale factor. Therefore, we leave the inclusion of radiation, matter and curvature for future work.

\section{Summary and conclusions}

In this paper, we have explored:{\it{``the existence of a fixed point in the effective average Lagrangian of asymptotic
safety quantum gravity, where a bounce arises naturally in the model by violating the NEC, and its consequences in the cosmology''}}.
To arrive at a meaningful implementation of the idea we had to define 
a mathematical procedure that transports the concept of a fixed point to a system of dynamical equations. 
First,
we defined and motivated the notion of an effective Lagrangian fixed point. Based on the hypothesis that such a fixed point exists in some regime of a scale-dependent theory gravitational theory, we formulated a model that implements such a ELFP in the low curvature expansion (\ref{eq_SkSO}).
Then we solved and analyzed the resulting field equations for a cosmology~(\ref{eq_ds2}). The solution has four integration constants $\Lambda_0, a_0, G_i, t_b$ and one parameter $c_L$, which we treat as the five-dimensional parameter space. 

In almost the entire parameter space, the solution naturally has a bouncing behavior for the scale factor $a(t)$. The influence of other dynamical variables on the parameter space is subtle. We have explored this aspect of our model by studying the time behavior of the gravitational coupling $G(t)$, curvature invariants, and the scale factor.
For $G(t)$, we find that beyond $t_{b,i}$, its evolution is continuous and converges asymptotically to zero. However, there is a notable exception in a four-dimensional subspace defined by $c_L = 2 \Lambda_0 / G_i$, where the gravitational coupling enters a regime of 'meta-stability' that delays its otherwise rapid decline to zero.

The scale factor and the derived Kretschmann scalar exhibit remarkable features. Notably, at times significantly later than $t_b$, the scale factor and curvature invariants mimic the behavior expected from classical Einstein equations with a cosmological constant and constant gravitational coupling, despite the actual variability and eventual finiteness of these couplings.


\section{Outlook}
In this work, we have provided a proof-of-principle demonstrating that it is possible to cast the concept of a fixed point into the form of dynamic equations. Moreover, the solutions to these equations incorporate a cosmological bounce in their dynamics.

In future efforts, it would be intriguing to investigate whether this conceptual toy model could be developed further into a viable model of the early universe. Specifically, we plan to explore the implications of relaxing the condition $T_{\mu \nu}=0$, or the assumption of spatial homogeneity."

\section*{Acknowlegements}

This work has been partially funded by Agencia Nacional de Investigaci\'{o}n y Desarrollo (ANID) through FONDECYT grant 1230112. A.R. is funded by the Generalitat Valenciana (Prometeo excellence programme grant CIPROM/2022/13) and by
the Maria Zambrano contract ZAMBRANO 21-25 (Spain).

\bibliography{ref}
\bibliographystyle{unsrt}

\end{document}